\documentclass[letterpaper,showpacs,superscriptaddress,twocolumn,aps,pra]{revtex4-1}
\usepackage{graphicx}

\begin{document}

\title{Fault-tolerant thresholds for quantum error correction with the surface code}
\author{Ashley M.~Stephens}\email{astephens@nii.ac.jp}
\affiliation{National Institute of Informatics, 2-1-2 Hitotsubashi, Chiyoda-ku, Tokyo 101-8430, Japan}

\date{\today}
\begin{abstract}
The surface code is a promising candidate for fault-tolerant quantum computation, achieving a high threshold error rate with nearest-neighbor gates in two spatial dimensions. Here, through a series of numerical simulations, we investigate how the precise value of the threshold depends on the noise model, measurement circuits, and decoding algorithm. We observe thresholds between 0.502(1)\% and 1.140(1)\% per gate, values which are generally lower than previous estimates.
\end{abstract}

\maketitle

\section{Introduction}
In theory, scalable quantum computation is possible if errors affecting qubits are not too strongly correlated and occur with a probability below some threshold value \cite{Aliferis06}. If the physical error rate is below the threshold, then quantum gates protected by an error-correction code can be arranged in a fault-tolerant manner such that any quantum circuit can be efficiently simulated to any accuracy \cite{Aharonov99,Kitaev97,Knill97,Preskill98}. The precise value of the threshold depends on an interplay between the effective noise in the quantum computer and the structure of the error-correction code in question, as well as the sophistication of the classical processing that accompanies the system \cite{Gottesman1}.

Recently, the surface code has emerged as a promising candidate for fault-tolerant quantum computation \cite{Kitaev2003,Bravyi2,Freedman1,Dennis2002,Raussendorf4,Raussendorf3,Raussendorf2007,Fowler1, Fowler2}. The surface code requires nearest-neighbor gates in two spatial dimensions with physical error rates of roughly one per cent or less, depending on the noise model. These requirements compare favorably with other codes, which may require non local gates \cite{Knill1} or may have significantly lower tolerance to errors \cite{Svore07,Stephens08,Stephens09,Spedalieri09}. For this reason, the surface code has underpinned several proposals for quantum computer architectures in a range of physical systems, including superconducting systems, atom-optical systems, trapped ions, quantum dots, and nitrogen-vacancy centers in diamond \cite{Fowler2,Stock09,Devitt09,Meter10,Yao12,Nickerson1,Monroe12,Nemoto13}.

This article concerns the value of the threshold error rate for the surface code. Previous numerical estimates of the threshold are in general agreement, ranging from 0.57\% to 1.40\% per gate, depending on various assumptions \cite{Raussendorf4, Raussendorf3,Raussendorf2007,Fowler1,Barrett1,Wang10,Wang11,Fowler3,Fowler2}. However, the use of different methods to arrive at these values makes it difficult to faithfully compare them. The threshold is an important target for experimental devices and, in part, determines the overhead of scalable quantum computation \cite{Devitt13}. Given this and considering the increasing relevance of the surface code to the development of quantum computer architectures, it is important to clearly understand how the precise value of the threshold depends on assumptions related to the noise model, measurement circuits, and decoding algorithm.

Here, through a series of numerical simulations, we investigate how the threshold is affected by these assumptions. We estimate thresholds for several syndrome measurement circuits under a range of physically motivated noise models. In general, our results highlight the dependency of the threshold on properties of the underlying physical system. In some cases, our results indicate that the threshold may be significantly lower than previously thought. Our work complements other recent results concerning the dependency of the threshold on correlated errors caused by the presence of a bosonic bath \cite{Mucciolo1,Mucciolo2} and on the effective noise in superconducting quantum circuits \cite{Ghosh1}.

Notwithstanding the recent development of several alternative decoding algorithms for topological codes \cite{Duclos-Cianci1,Duclos-Cianci2,Bravyi1b,Sarvepalli1,Wootton1,Wootton2,Wootton3,Delfosse1}, we restrict ourselves to decoding via Edmonds' minimum-weight perfect matching algorithm \cite{Edmonds1}. Also, we do not consider other topological codes, such as color codes \cite{Bombin1}, instead, referring the interested reader to the recent article of Landahl {\it et al.~}\cite{Landahl1}.
\begin{figure}
\begin{center}
\includegraphics[scale=0.35]{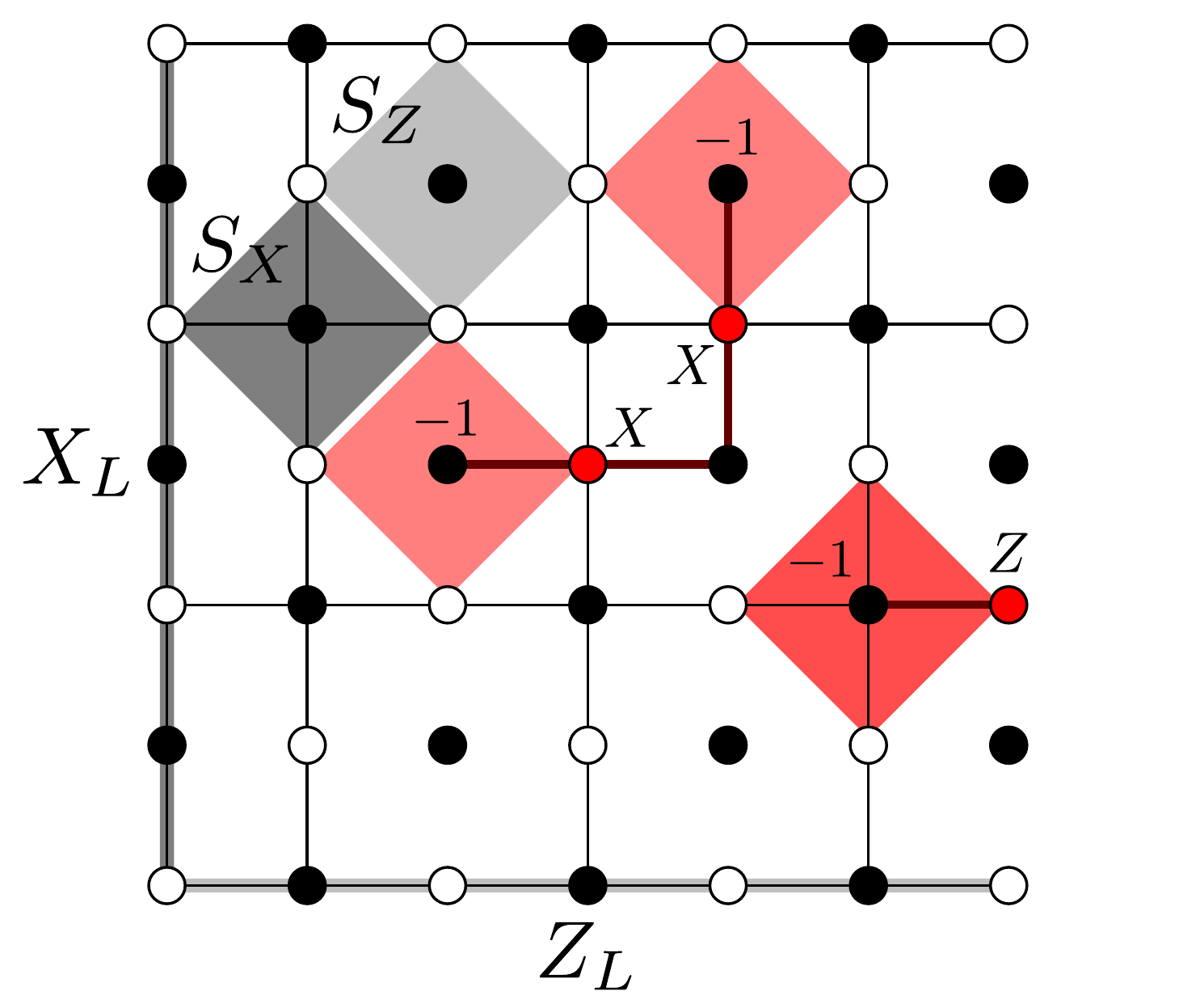}
\end{center}
\vspace{-17pt}
\caption{(Color online) Structure of the surface code for $d=4$ where open circles signify data qubits and closed circles signify ancillary qubits. Stabilizer generators and logical operators are indicated. Chains of $X$ and $Z$ errors affecting data qubits will anticommute with the stabilizer generators at the endpoints, which will have eigenvalues equal to $-1$ as indicated. End points may be obscured if chains terminate on boundaries. In general, the number of data qubits is $d^2+(d-1)^2$.} 
\label{surface}
\end{figure}
\begin{figure}
\begin{center}
\includegraphics[scale=1.00]{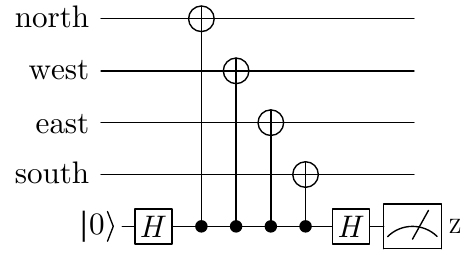}
\end{center}
\vspace{-12pt}
\caption{Circuit to measure the eigenvalue of the stabilizer generator $S_X$ associated with a vertex where the order of operations is defined in relation to the ancillary qubit at that vertex. The circuit depth is eight.}
\label{sx}
\end{figure}
\begin{figure}
\begin{center}
\includegraphics[scale=1.00]{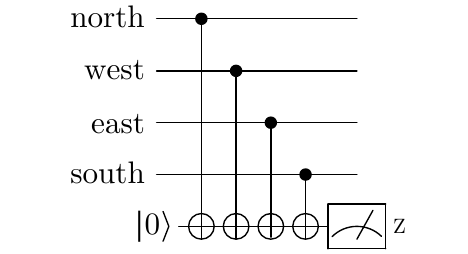}
\end{center}
\vspace{-12pt}
\caption{Circuit to measure the eigenvalue of the stabilizer generator $S_Z$ associated with a face. The circuit depth is six.}
\label{sz}
\end{figure}

\section{The surface code}
The surface code, also known as the planar code, is a variation of Kitaev's toric code \cite{Kitaev2003}. The toric code is defined over $2d^2$ qubits located on the edges of a $d\times d$ square lattice embedded on a two-dimensional torus, where $d$ is the code distance. The four-dimensional code space is the simultaneous +1 eigenspace of the stabilizer generators \cite{Gottesman97}, defined as 
\begin{equation}
S_X=\bigotimes_{i\in n(v)}X_i,
\end{equation}
and 
\begin{equation}
S_Z= \bigotimes_{j\in n(f)}Z_j,
\end{equation}
where $v$ is a vertex in the embedding, $f$ is a face in the embedding, $n$ refers to the four neighboring qubits, and $X$ and $Z$ are the usual single-qubit Pauli operators. The surface code is similarly defined, but its topology is modified from a torus to a two-dimensional plane with boundaries that alternate between open and closed faces. Then, the two-dimensional code space encodes a single logical qubit \cite{Bravyi2,Freedman1}. The logical Pauli operators are the pair of homologically nontrivial chains of $X$ and $Z$ operators that connect opposite boundaries of the same kind, which preserve the code space, as they commute with the stabilizer generators but act nontrivially on the logical qubit. Although the logical Pauli operators can be deformed by the stabilizer generators, their minimum length is always equal to $d$. The structure of the surface code for $d=4$ is illustrated in Fig.~\ref{surface}. 

Universal quantum computation is achieved by manipulating the logical operators, using the techniques developed by Raussendorf {\it et al.~}\cite{Raussendorf4,Raussendorf3}. By defining a surface code on a plane with a more complicated topology, multiple logical qubits are introduced. The various logical operators are manipulated by deforming the topology of the surface through a series of measurements \cite{Fowler1}. Here, we restrict our study to the case where a surface code encodes a single logical qubit. In particular, we are interested in the active process of quantum error correction, which is used to preserve the quantum information stored in the surface code. Since this process is largely unchanged in the presence of additional logical qubits, our results are applicable in general.

\section{Measuring and interpreting \\the error syndrome}
Pauli errors affecting qubits in the surface code anticommute with a subset of the stabilizer generators. For example, an $X$ error anticommutes with the $Z$-type stabilizer generators associated with the adjacent vertices, which will have eigenvalues equal to $-1$. Connected chains of errors anticommute with the stabilizer generators at the end points of the chains, which may be hidden if the chains terminate on boundaries as shown in Fig.~\ref{surface}.

In order to identify errors, we measure the eigenvalues of the stabilizer generators, giving us an error syndrome. These measurements are performed by introducing ancillary qubits as shown in Fig.~\ref{surface} and executing the measurement circuits shown in Figs.~\ref{sx} and \ref{sz}. The circuits require nearest-neighbor gates in two spatial dimensions and can be performed in parallel (with one circuit for each stabilizer generator) across the entire surface code. In general, the error syndrome may be unreliable due to errors affecting the ancillary qubits, such as measurement errors. To mitigate this, the measurement circuits are repeated $d$ times, and we record when a measurement outcome changes from its previous value, which indicates that an error of some kind has occurred. An error affecting a data qubit will cause a pair of measurements separated in space to change from its previous values, whereas, an error affecting an ancillary qubit will cause a single measurement to change from its previous value and then to immediately change back again. In general, connected chains of errors can involve both kinds of errors, so the end points, indicated by the changing measurement outcomes, may be separated in both space and time. Thus, the error syndrome is the entire space-time volume of these changes.

Since errors perturb the state of the system from the code space, error correction involves identifying a set of corrections that will restore the state to the code space while preserving the encoded quantum information. There are several algorithms to interpret or to decode the error syndrome, which, in general terms, balance accuracy (having a high likelihood of identifying the correct homology class of the errors) with efficiency (capable of decoding the syndrome for large codes in a sufficiently short time) \cite{Dennis2002,Wang03,Harrington04,Duclos-Cianci1,Duclos-Cianci2,Bravyi1b,Sarvepalli1,Wootton1,Wootton2,Wootton3,Delfosse1}. Here, we use a decoding algorithm that identifies the most likely set of errors consistent with the error syndrome where we consider $X$ and $Z$ errors separately \cite{Dennis2002,Raussendorf3,Wang10}. In the algorithm, each measurement change is represented by a node in a graph. Edges between nodes are weighted to reflect the probability of the associated measurement changes being caused by a connected chain of errors. A perfect matching of the graph reveals a set of errors consistent with the error syndrome, and the minimum-weight perfect matching reveals the most likely set. From this set, an appropriate correction can be inferred. Care must be taken to account for correlated errors that arise in the measurement circuits, and edges should be appropriately weighted to account for the fact that different kinds of errors (which cause different pairs of measurement changes) may occur with different probabilities \cite{Raussendorf3}.

\section{Overview of numerical methods}
Our aim is to determine the threshold error rate of the surface code. For physical error rates below this value, increasing the code distance (linearly) will decrease the logical error rate (exponentially). To determine the logical error rate as a function of the physical error rate, we perform Monte Carlo simulations. In each instance, a set of errors is generated based on some noise model, the error syndrome is calculated and decoded, a correction is applied, and the resulting homology class is calculated to test for the presence or absence of a logical error. For noise models in which the error syndrome is unreliable, the measurement circuits are repeated $d$ times before the error syndrome is decoded. In our simulations, minimum-weight perfect matching is performed with Kolmogorov's implementation \cite{Kolmogorov1} of Edmonds' perfect matching algorithm \cite{Edmonds1}, and we use a Mersenne twister pseudo-random number generator \cite{Saito1}. For each physical error rate, the logical error rate is an average of approximately $10^6$ independent instances, where we ensure that at least $10^4$ logical errors are observed per point to limit statistical uncertainty.

For a local error model, decoding of the surface code can be mapped to a three-dimensional random-plaquette gauge model on classical spins where the zero-temperature phase transition corresponds to the threshold error rate \cite{Dennis2002,Wang03,Harrington04}. Following Wang {\it et al.~}\cite{Wang03}, the behavior of the logical error rate near the threshold corresponds to critical behavior in the spin model where the spin-correlation length $\xi$ scales according to 
\begin{equation}
\xi\sim\vert p-p_{th} \vert^{-\nu_0},
\end{equation}
where $p$ is some physical error rate, $p_{th}$ is the threshold error rate, and $\nu_0$ is the scaling exponent corresponding to the universality class of the model. Thus, for sufficiently large $d$, the logical error rate $p_l$ should follow
\begin{equation}
p_l=(p-p_{th})d^{1/\nu_0}.
\end{equation}
Allowing for systematic finite-size effects, we fit our data to a quadratic universal scaling function,
\begin{equation}
p_l=A+B(p-p_{th})d^{1/\nu_0}+C(p-p_{th})^2d^{2/\nu_0},
\label{scaling}
\end{equation}
from which we determine $p_{th}$ and $\nu_0$. We perform simulations for odd values of $d$ between $d=3$ and $d=17$ where $p\sim p_{th}$. Violations of the scaling ansatz are discernible for the smallest codes such that the minimum code distance for strong agreement between the numerical data and the ansatz appears to be $d=7$. To account for this, values of $p_{th}$ and $\nu_0$ are determined from a best fit of the data for $d\geq9$. In every case, $R^2>0.999$, indicating accurate fitting. When plotting the data in Figs.~\ref{std} and \ref{bal}, the curves for $d\geq9$ follow the universal scaling function in Eq.~(\ref{scaling}). Data for $d\leq7$ are included for completeness, however, the corresponding curves are independent polynomial fits that serve only as a guide for the eye. Our results indicate that, for the various circuit-based noise models we consider, which introduce only short-range correlated errors, the value of $\nu_0$ is consistent with the universality class of the strictly local three-dimensional random-plaquette gauge model \cite{Wang03}.

The surface code is defined by its hard boundaries. However, it has been common to, instead, study the threshold of the toric code, which effectively has periodic boundary conditions in two spatial dimensions. Here, we present results for the surface code. In this case, the measurement circuits at the boundaries of the surface code are modified to account for the omitted qubits. This changes their effective error rate from the measurement circuits in the bulk. However, we will see that the logical error rate rapidly converges to a single value at the threshold as the code distance is increased, indicating that these boundary effects are significant only for the smallest codes. This suggests that the toric code and the surface code will share the same threshold. However, because the structure of the logical operators depends on the boundary conditions, the correct boundary conditions should be used when an estimate of the logical error rate is sought for some physical error rate.

Lastly, the threshold is sensitive to errors that arise in the measurement circuits, which will, in turn, depend on the set of gates native to the quantum computer. We consider three cases, which are parametrized by the overall circuit depth:
\begin{enumerate}
\item {\it Depth-eight circuits.} First, we assume the gate set consists of the preparation of state $\vert0\rangle$, the single-qubit Hadamard rotation, the two-qubit controlled-\textsc{not} gate, and measurement in the $Z$ basis. Then, referring to the circuits in Figs.~\ref{sx} and \ref{sz}, the overall circuit depth is equal to eight. In this case, there is an asymmetry between the two measurement circuits with the longer circuit being more unreliable due to the additional gates. This causes the threshold to split into an $X$-error threshold and a $Z$-error threshold.

\item {\it Depth-six circuits.} Second, we assume that the gate set is extended to include the preparation of state $\vert+\rangle$ and measurement in the $X$ basis. This removes the need for the Hadamard rotations in Fig.~\ref{sx}, and so, the overall circuit depth is reduced to six.

\item {\it Depth-five circuits.} Third, we assume measurement is nondestructive and prepares the ancillary qubit in a known state (either $\vert+\rangle$ and $\vert-\rangle$ or $\vert0\rangle$ and $\vert1\rangle$, depending on the measurement basis). This allows the measurement and state preparation to be combined, and so, the overall circuit depth is reduced to five. 
\end{enumerate}

In each of these cases, all measurement circuits are performed in parallel and repeated $d$ times where identity gates are inserted whenever qubits are required to be idle. In the first case, we give the $Z$-error threshold, which is the lower of the two thresholds and, therefore, sets the overall threshold. In all other cases, we give the $X$-error threshold. These thresholds set targets for the high-level gates specified in the circuits, rather than for any lower-level physical operations. Also, we have ignored gates that are not required for error correction, but which may be required to achieve universality by distillation \cite{BK05}.

\section{Numerical results}
\subsection{Code capacity noise model} We begin with an idealized case in which the error syndrome of the surface code can be measured perfectly. Single-qubit Pauli errors are applied to data qubits with probability $p$. In this case, we are effectively testing the code capacity of the surface code. Because it is perfectly reliable, the error syndrome only needs to be measured once. This eliminates the timelike aspect of the decoding algorithm, and error correction is reduced to interpreting the error syndrome in two spatial dimensions. Note that this simplified decoding problem can be mapped to the two-dimensional random-bond Ising model on classical spins \cite{Dennis2002,Wang03,Harrington04,Stace1}. For the code capacity noise model, we find
\begin{eqnarray}
p_{th}&=& 0.1030\pm0.0001,\\
\nu_0&=&1.47\pm 0.01,
\end{eqnarray}
consistent with Wang {\it et al.~}\cite{Wang03}, who found $p_{th}=0.1031\pm0.0001$ and $\nu_0=1.46\pm0.01$. Our threshold is lower than the threshold of $\sim$0.109 found for an optimal decoding algorithm \cite{Honecker,Merz,Ohzeki09,deQueiroz09} but higher than the threshold of $\sim$0.09 found for a renormalization-group decoding algorithm \cite{Duclos-Cianci1}. 

\subsection{Phenomenological noise model}
Next, we move to a case in which errors can occur on both data and ancillary qubits. Single-qubit Pauli errors are applied to all qubits with probability $p$. This noise model neglects the propagation of errors between data and ancillary qubits in the measurement circuits but captures the essential challenge of fault-tolerant error correction where the process of error correction itself is inherently faulty. In this case, the full decoding algorithm is required to account for the unreliable error syndrome. For the phenomenological noise model, we find
\begin{eqnarray}
p_{th}&=& 0.0290\pm 0.0001,\\
\nu_0&=&1.01\pm 0.01.
\end{eqnarray}
Again, this is consistent with Wang {\it et al.~}\cite{Wang03}, who found $p_{th}=0.0293\pm0.0002$ and $\nu_0=1.00\pm0.05$. Our threshold is lower than the threshold of $\sim$0.033 found for an optimal decoding algorithm \cite{Ohno04} but higher than the threshold of $\sim$0.0194 found for a renormalization-group decoding algorithm \cite{Duclos-Cianci2}. 

\begin{figure*}
\begin{center}
\includegraphics[scale=0.465]{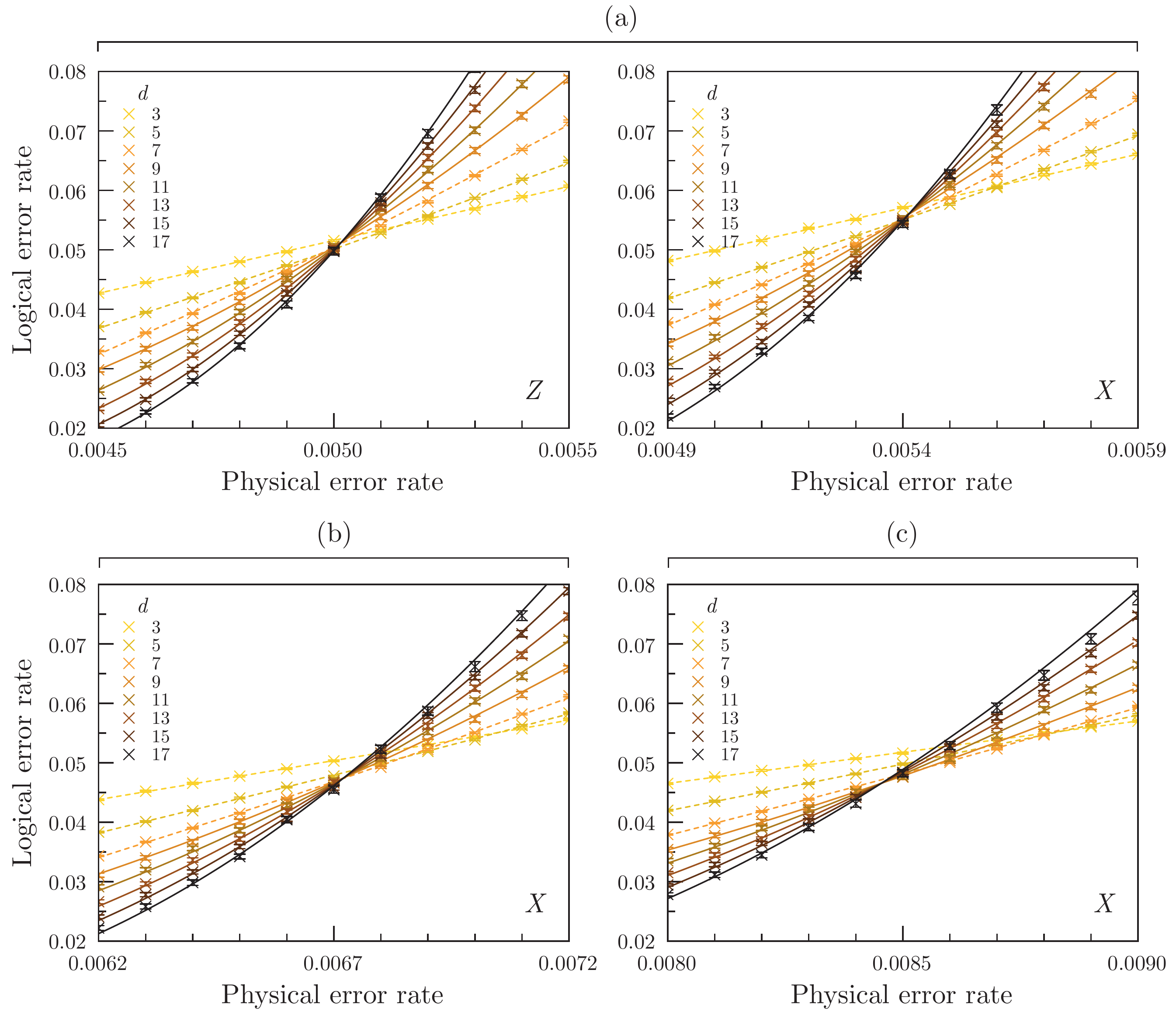}
\end{center}
\vspace{-15pt}
\caption{(Color online) Logical error rate as a function of the physical error rate for the standard circuit-based noise model for various code distances. Solid curves are derived from the universal scaling function in Eq.~(\ref{scaling}), and dashed curves are polynomial fits that serve only as a guide for the eye. Error bars indicate a $\pm2\sigma$ statistical error. The value of the physical error rate at the intersection is the threshold. (a) Depth-eight circuits where the threshold splits into an $X$-error threshold and a $Z$-error threshold due to the asymmetry between the two measurement circuits. The $Z$-error threshold is lower and, therefore, sets the overall threshold. (b) Depth-six circuits. (c) Depth-five circuits.}
\label{std}
\end{figure*}

\begin{figure*}
\begin{center}
\includegraphics[scale=0.465]{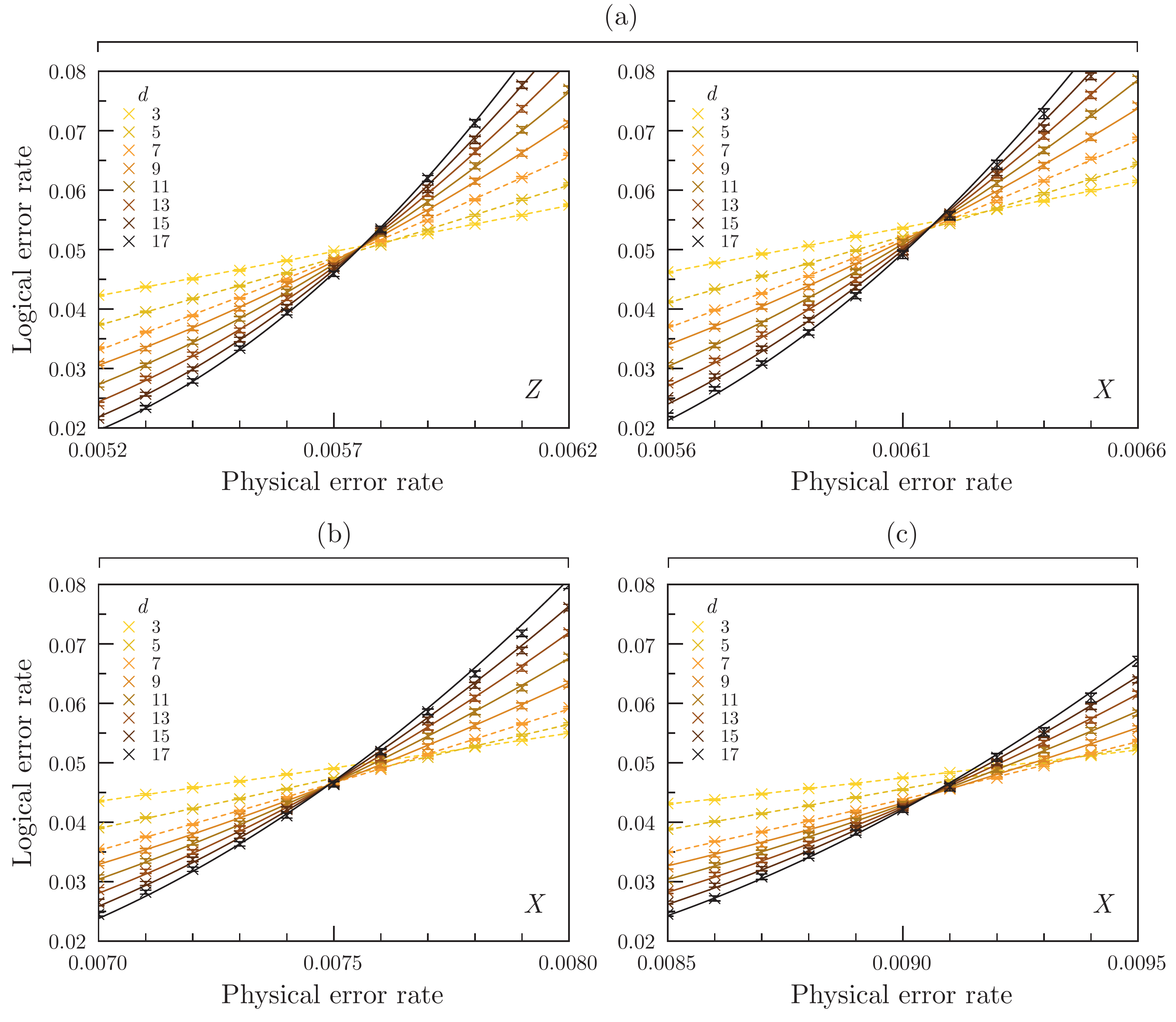}
\end{center}
\vspace{-15pt}
\caption{(Color online) Logical error rate as a function of the physical error rate for the balanced circuit-based noise model for various code distances. See the caption for Fig.~\ref{std} for details. (a) Depth-eight circuits. (b) Depth-six circuits. (c) Depth-five circuits.} 
\label{bal}
\end{figure*}

\subsection{Standard circuit-based noise model}
Next, we move to a more general noise model, assuming that all gates in the measurement circuits may introduce errors. This is the most relevant case for fault-tolerant quantum computation, although we note that the particulars of the noise model will depend on the physical system under consideration. For example, measurements may be slower and less reliable than other gates. First, we consider a so-called standard noise model. Erroneous single-qubit gates occur with probability $p$, acting ideally followed by a single-qubit Pauli error chosen randomly from set $\{X,Y,Z\}$. Similarly, erroneous two-qubit gates occur with probability $p$, acting ideally followed by a two-qubit Pauli error chosen randomly from set $\{IX,IY,IZ,XI,\dots,ZZ\}$. Lastly, erroneous initialization and measurement each occur with probability $p$, preparing or reporting the incorrect orthogonal eigenstate. Under the standard noise model, for the depth-eight circuits, we find
\begin{eqnarray}
p_{th}&=& 0.00502\pm0.00001,\\
\nu_0&=&1.05\pm 0.01,
\end{eqnarray}
for the depth-six circuits, we find
\begin{eqnarray}
p_{th}&=& 0.00672\pm0.00001,\\
\nu_0&=&1.06\pm 0.02,
\end{eqnarray}
and, for the depth-five circuits, we find
\begin{eqnarray}
p_{th}&=& 0.00846\pm0.00001,\\
\nu_0&=&1.02\pm 0.01,
\end{eqnarray}
as shown in Fig.~\ref{std}.

\subsection{Balanced circuit-based noise model}
The standard noise model is somewhat unreasonable as the qubits involved in two-qubit gates are more reliable than idle qubits. So, next, we consider a so-called balanced noise model, which ensures that idle qubits have the same probability of error as the qubits involved in two-qubit gates and accounts for the fact that measurement is only sensitive to errors in one basis. Specifically, the standard noise model is modified so that erroneous single-qubit gates occur with the probability of $4p/5$ and erroneous initialization and measurement occurs with the probability of $2p/3$. Under the balanced noise model, for the depth-eight circuits, we find
\begin{eqnarray}
p_{th}&=& 0.00576\pm0.00001,\\
\nu_0&=&1.05\pm 0.02,
\end{eqnarray}
for the depth-six circuits, we find
\begin{eqnarray}
p_{th}&=& 0.00749\pm0.00001,\\
\nu_0&=&1.02\pm 0.01,
\end{eqnarray}
and, for the depth-five circuits, we find
\begin{eqnarray}
p_{th}&=& 0.00905\pm0.00001,\\
\nu_0&=&1.00\pm 0.01,
\end{eqnarray}
as shown in Fig.~\ref{bal}.

\subsection{Perfect single-qubit gates}
In some physical systems, single-qubit gates may be significantly faster and more reliable than two-qubit gates. In this case, the threshold will depend mainly on the two-qubit controlled-\textsc{not} gates in the measurement circuits. We can approximate this case by modifying the standard noise model so that all single-qubit gates (including measurement and initialization) are perfectly reliable. In this case, we find
\begin{eqnarray}
p_{th}&=& 0.01140\pm0.00001,\\
\nu_0&=&1.05\pm 0.03.
\end{eqnarray}

\subsection{Decoding algorithm with a rectilinear metric}
Next, we consider the effect of simplifying the decoding algorithm. Following Raussendorf {\it et al.~}\cite{Raussendorf3}, our decoding algorithm accounts for the relative probabilities of errors, including correlated errors, that arise in the measurement circuits. However, the threshold was previously estimated using a decoding algorithm that ignores these correlated errors \cite{Fowler1,Wang10}. This algorithm is also based on minimum-weight matching on a graph, but the weights of edges between nodes are made to equal the rectilinear distance between those nodes, simply reflecting the minimum number of single-qubit Pauli errors in a chain connecting the endpoints. Without accounting for correlated errors, the surface code corrects fewer errors than the code distance implies, negatively affecting its performance, particularly at low error rates. In fact, for $d=3$, the code cannot reliably correct even a single error. With this simplification, under the standard noise model, for the depth-six circuits, we find
\begin{eqnarray}
p_{th}&=& 0.00504\pm0.00001,\\
\nu_0&=&0.99\pm 0.02.
\end{eqnarray}
Fortunately, there is no significant cost to accurately accounting for correlated errors in the surface code. Similar methods exist for accounting for correlated errors in concatenated quantum error correction, also leading to significantly improved performance \cite{Knill1,Poulin1,Evans2}.

\subsection{Three-dimensional topological cluster states}
Lastly, we consider an interesting and closely related scheme known as topological cluster-state quantum error correction \cite{Raussendorf4,Raussendorf3}. In this scheme, the measurement circuits are simulated by a series of single-qubit measurements on a particular three-dimensional cluster state \cite{Raussendorf3}. The scheme may be more practical than the surface code in some physical systems partly due to its elegant tolerance against qubit loss, which was shown by Barrett and Stace \cite{Barrett1}. A modified depth-six circuit is required to prepare the cluster state from unentangled qubits and then to measure each qubit in the appropriate basis \cite{Raussendorf3}. However, the decoding algorithm is largely unchanged from the algorithm for the surface code. Under the standard noise model, we find
\begin{eqnarray}
p_{th}&=& 0.00545\pm0.00001,\\
\nu_0&=&1.01\pm0.01,
\end{eqnarray}
and, under the balanced noise model, we find
\begin{eqnarray}
p_{th}&=& 0.00626\pm0.00001,\\
\nu_0&=&1.01\pm0.01.
\end{eqnarray}

\begin{table*}
\begin{center}
\vspace*{4pt}   
\begin{tabular}{lll}
\hline\hline
\hspace{4.0cm} & Standard noise model \hspace{0.5cm} & Balanced noise model \\
\hline
Depth-eight circuits & $0.00502\pm0.00001$ & $0.00576\pm0.00001$ \\
& $0.00541\pm0.00001$\footnote{Threshold for $X$ errors. For this noise model, the $Z$-error threshold is lower and, therefore, sets the overall threshold.} & $0.00616\pm0.00001$\footnotemark[1] \\
& 0.0057\footnotemark[1]\footnote{Estimated from the logical error rate per round of measurement, rather than per $d$ rounds of measurement.} \cite{Fowler2} & \\
\hline
Depth-six circuits & $0.00672\pm0.00001$ & $0.00749\pm0.00001$ \\
& & 0.0075\footnote{Not directly comparable due to minor differences in the measurement circuits and noise model.} \cite{Raussendorf3} \\
\hline
Depth-five circuits & $0.00846\pm0.00001$ & $0.00905\pm0.00001$ \\
& 0.009\footnotemark[2] \cite{Fowler3} & 0.0012\footnotemark[2] \cite{Wang11} \\
& 0.011\footnotemark[2] \cite{Wang11} & \\
\hline
Perfect single-qubit gates & $0.01140\pm0.00001$ & -- \\
& 0.0125\footnotemark[2] \cite{Fowler2} & \\
& 0.014\footnotemark[2] \cite{Wang11} & \\
\hline
Rectilinear metric & $0.00504\pm0.00001$ & -- \\
(depth-six circuits) & 0.006\footnotemark[2] \cite{Fowler1} & \\
& 0.0078\footnotemark[2] \cite{Wang10} & \\
\hline
Topological cluster states & $0.00545\pm0.00001$ & $0.00626\pm0.00001$ \\
& & 0.0063 \cite{Barrett1} \\
& & 0.0067\footnotemark[3] \cite{Raussendorf4} \\
\hline\hline
\end{tabular}
\vspace{7pt}
\caption{Summary of thresholds for various circuit-based noise models where our results are compared with previous estimates. The values without references are our results. If no uncertainty is given, none was reported in the associated reference.}
\label{thr}
\end{center}
\end{table*}

\section{Discussion}
It is instructive to compare our results with a range of previous estimates of the threshold. We begin by noting that it is reasonable to expect some slight variation between estimates due to different implementations of the decoding algorithm and the numerical simulations. Nevertheless, for the code capacity and phenomenological noise models, our results are consistent with Wang {\it et al.~}\cite{Wang03}. For the remaining circuit-based noise models, our results are summarized in Table \ref{thr} and are compared with a range of previous estimates. Of the values that can be directly compared, our results are consistent only with the estimate of the threshold for topological cluster-state error correction due to Barrett and Stace \cite{Barrett1}. Beyond this result, there is some variation with our thresholds being significantly lower than those previously reported. This discrepancy appears to be independent of the particular measurement circuit, noise model, and decoding algorithm.

To investigate this discrepancy, let us consider the definition of the logical error rate. Recall that measurement circuits are repeated to account for the fact that the error syndrome is unreliable. We define the logical error rate to be the error rate per $d$ rounds of measurement, following Raussendorf {\it et al.~}\cite{Raussendorf4,Raussendorf3}. This definition reflects the fact that, for a roughly isotropic noise model, $d$ rounds are required to achieve the same protection against errors affecting ancillary qubits as against errors affecting data qubits. In other words, if we increase the code distance, then error correction takes more time, which should be accounted for when calculating the logical error rate. On the other hand, the estimates in Refs.~\cite{Fowler1,Wang10,Wang11,Fowler3,Fowler2} share a different definition (also see Refs.~\cite{Nickerson1,Bravyi13}). According to this definition, the logical error rate is the error rate per round of measurement (or, equivalently, the logical error rate per round is reciprocated to give the expected number of rounds until a logical error occurs). Note that this definition is independent of the code distance $d$. In both cases, for various code distances, the logical error rate is plotted over a range of physical error rates, and the threshold is estimated to be the physical error rate for which these curves intersect.
\begin{figure}
\begin{center}
\includegraphics[scale=0.465]{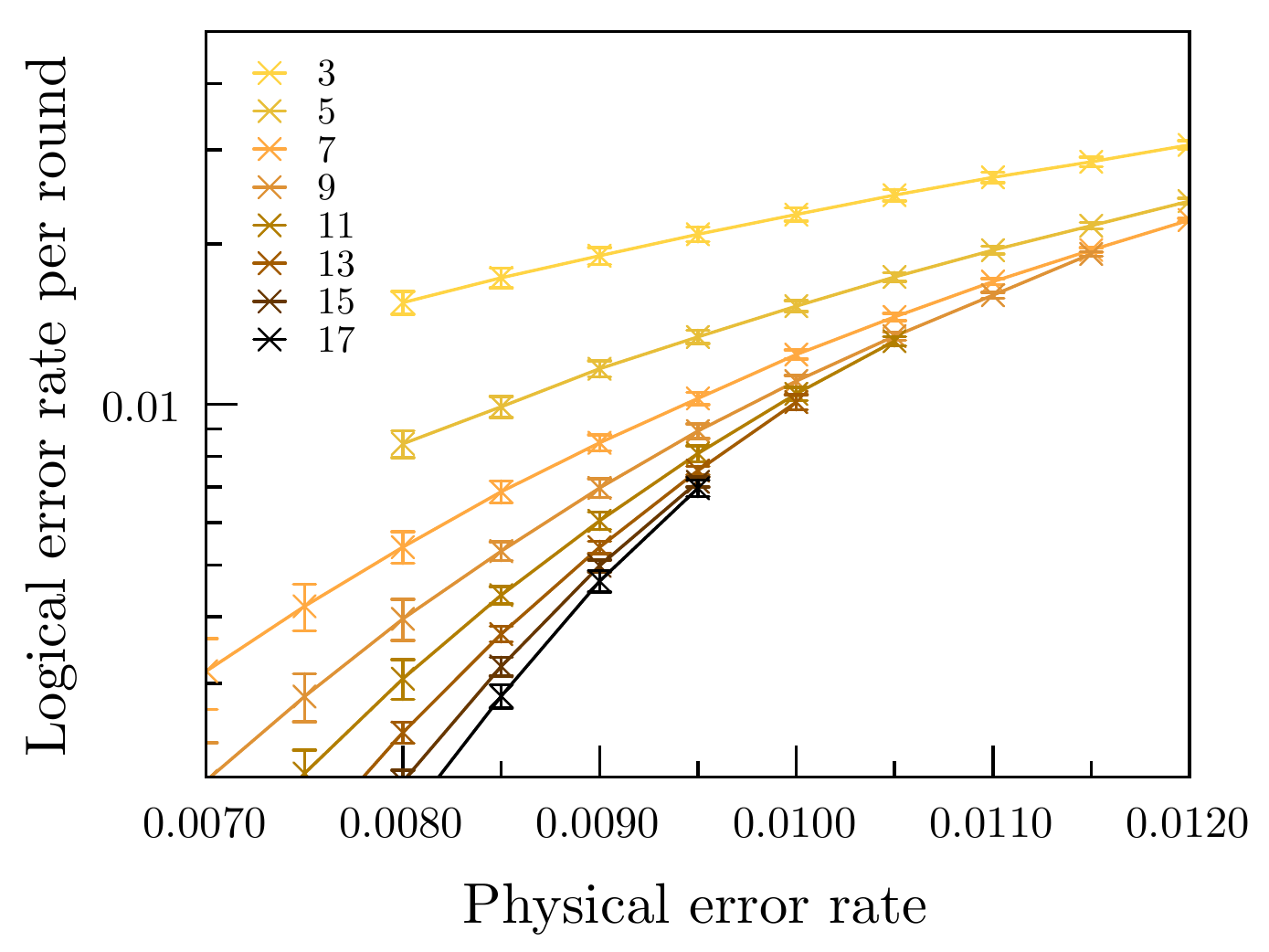}
\end{center}
\vspace{-15pt}
\caption{(Color online) Logical error rate per round of measurement as a function of the physical error rate for the depth-five circuits under the standard noise model. Error bars indicate a $\pm2\sigma$ statistical error. [Compare with Fig.~\ref{std}(c).]} 
\label{ttf}
\end{figure}

Let us define the logical error rate to be the error rate per round of measurement as per Refs.~\cite{Fowler1,Wang10,Wang11,Fowler3,Fowler2} and consider two surface codes with code distances $d=n$ and $d=n+2$. For some physical error rate $p_n$, the logical error rate of the two codes will be equal. However, if we fix the physical error rate at $p_n$ and perform $d$ rounds of measurement as required, then the larger surface code will be more likely to fail. In other words, according to this new definition, the two codes are equally reliable, but according to our original definition, the larger code is less reliable. The latter implies that the threshold is actually at some physical error rate $p_{th}<p_n$. If $n$ becomes larger, then the relative difference between the two code distances becomes smaller as does the relative difference between their reliability over $d$ rounds of measurement. So, as $n\rightarrow\infty$, we may expect $p_n\rightarrow p_{th}$ from above. This would suggest that defining the logical error rate to be the error rate per round of measurement could lead to an overestimate of the threshold.

To test this assertion, we return to the depth-five circuits under the standard noise model. Figure \ref{ttf} shows the logical error rate per round of measurement as a function of the physical error rate for various code distances. As the code distance increases, the physical error rate at which consecutive curves intersect decreases. In particular, the intersection moves from physical error rates above 0.01 to approximately 0.0095. This is roughly consistent with the data in Refs.~\cite{Wang11,Fowler3}, and qualitatively similar behavior can also be seen in Refs.~\cite{Fowler1,Wang10,Fowler2}. In Ref.~\cite{Wang11}, the threshold was estimated to be $0.011$ from the intersection of the $d=7$ and $d=9$ curves, and in Ref.~\cite{Fowler3}, the threshold was estimated to be $0.009$ from the intersection of the $d=45$ and $d=55$ curves. The discrepancy between these two values was attributed to significant boundary effects for $d\leq21$. However, our earlier simulations indicate that boundary effects are negligible for $d\geq7$, pointing to another explanation for this behavior. Also, in Ref.~\cite{Fowler3}, there appears to be no consistent intersection, even for $d>25$. This implies that the threshold is actually lower than 0.009. Recall that, under the same assumptions, we found $p_{th}=0.00846\pm0.00001$ as shown in Fig.~\ref{std}(c). Ultimately, given the lack of error analysis in Refs.~\cite{Fowler1,Wang10,Wang11,Fowler3,Fowler2}, it is difficult to make a conclusive statement about the discrepancy between these estimates and our results.

\section{Conclusion and further work}
To summarize, we have performed a series of numerical simulations of the surface code, finding that the value of the threshold error rate varies between 0.502(1)\% and 1.140(1)\% per gate for typical assumptions made in studies of this kind. Our results highlight the dependency of the threshold on properties of the underlying physical system. For example, having to perform additional gates to access initialization and measurement in the conjugate basis significantly reduces the threshold. Similarly, the highest thresholds will only be realized if measurements (in both the $X$ and $Z$ bases) are nondestructive or if all single-qubit gates are effectively free from noise. However, in some cases, our results indicate that the threshold may be significantly lower than previously thought. The target for experimental devices may be lower still, assuming that gates, such as the two-qubit controlled-\textsc{not} gate, will be composed of several physical operations. The operational error rate must also be sufficiently below the threshold to limit the overhead due to error correction. Lastly, our results indicate that the threshold for topological cluster-state error correction is lower than for the surface code under an identical noise model. However, like other schemes based on cluster states, this scheme has several desirable properties that may offset this disadvantage in some physical systems, particularly systems with nondeterministic gates or systems significantly affected by qubit loss or leakage.

We have limited ourselves to the question of the threshold for the surface code with decoding via Edmonds' minimum-weight perfect matching algorithm. Naturally, there are many avenues for further work. Given the recent proliferation of alternative decoding algorithms for topological codes, such as the surface code \cite{Duclos-Cianci1,Duclos-Cianci2,Bravyi1b,Sarvepalli1,Wootton1,Wootton2,Wootton3,Delfosse1}, it would be valuable to determine circuit-level thresholds for these algorithms, making it easier to understand their practical costs and benefits. It may also be possible to improve these thresholds by accounting for additional correlations present in some noise models (for example, the correlation between $X$ and $Z$ errors in depolarizing noise) \cite{Duclos-Cianci1}. Comparing these thresholds in a consistent manner will be necessary to draw strong conclusions about the different approaches to error correction in the surface code.

Another important open question is the performance of the surface code at error rates well below the threshold. A greater understanding of this regime---including an understanding of how performance is affected by the introduction of additional logical qubits and nontrivial logical gates---will assist in determining the true overhead of scalable quantum computation under various assumptions. This question was recently addressed by Bravyi and Vargo for the standard noise model \cite{Bravyi13}. Expanding their work to consider a range of noise models and decoding algorithms would be instructive.

Lastly, we highlight related schemes for topological quantum error correction against noise models that differ significantly from the typical models considered here. These include schemes to tolerate high rates of qubit loss \cite{Barrett1,Barrett2, Fujii10} and a concatenated code tailored to highly dephasing-biased noise \cite{Stephens13}. Considering other physically motivated noise models may lead to new schemes that could underpin quantum computer architectures in the future.

{\it Acknowledgements---} This work was supported by the FIRST Program in Japan. Thanks to W.~Munro and K.~Nemoto for helpful advice and to M.~Carrasco for commenting on several versions of the paper.

\bibliographystyle{unsrt}

\begin{thebibliography}{10}

\bibitem{Aliferis06}
P.~Aliferis, D.~Gottesman, and J.~Preskill, {\it Quantum Inf. Comput.} \textbf{6}, 97 (2006).

\bibitem{Aharonov99}
D.~Aharonov and M.~Ben-Or, in {\it Proceedings of the 29th ACM Symposium on the Theory of Computation} (Association for Computing Machinery, New York, 1998), p. 176.

\bibitem{Kitaev97}
A.~Y.~Kitaev, {\it Russian Math. Surveys} {\bf 52}, 1191 (1997).

\bibitem{Knill97}
E.~Knill, R.~Laflamme, and W.~Zurek, {\it Proc. R. Soc. London, Ser.~A} {\bf 454}, 365 (1998).

\bibitem{Preskill98}
J.~Preskill, in {\it Introduction to Quantum Computation and Information}, edited by H.-K.~Lo, T.~Spiller, and S.~Popescu (World Scientific, Singapore, 1998).

\bibitem{Gottesman1}
D.~Gottesman. {\it Proc. Symp. Appl. Math.} {\bf 68}, 13 (2009).

\bibitem{Kitaev2003}
A.~Y.~Kitaev, {\it Ann. Phys.} {\bf 303}, 2 (2003).

\bibitem{Bravyi2}
S.~B.~Bravyi and A.~Y.~Kitaev, arXiv:quant-ph/9811052.

\bibitem{Freedman1}
M.~H.~Freedman and D.~A.~Meyer, {\it Found. Comp. Math.} {\bf 1}, 325 (2001).

\bibitem{Dennis2002}
E.~Dennis, A.~Kitaev, A.~Landahl, and J.~Preskill, {\it J. Math. Phys.} {\bf 43}, 4452 (2002).

\bibitem{Raussendorf4}
R.~Raussendorf, J.~Harrington, and K.~Goyal, {\it Ann. Phys.} {\bf 321}, 2242 (2006).

\bibitem{Raussendorf3}
R.~Raussendorf, J.~Harrington, and K.~Goyal, {\it New J. Phys.} {\bf 9}, 199 (2007).

\bibitem{Raussendorf2007}
R.~Raussendorf and J.~Harrington, {\it Phys. Rev. Lett.} {\bf 98}, 190504 (2007).
	 
\bibitem{Fowler1}
A.~G.~Fowler, A.~M.~Stephens, and P.~Groszkowski, {\it Phys. Rev. A} \textbf{80}, 052312 (2009).

\bibitem{Fowler2}
A.~G.~Fowler, M.~Mariantoni, J.~M.~Martinis, and A.~N.~Cleland, {\it Phys. Rev. A} \textbf{86}, 032324 (2012).

\bibitem{Knill1}
E.~Knill. {\it Nature (London)} \textbf{434}, 39 (2005).

\bibitem{Svore07}
K.~M.~Svore, D.~P.~DiVincenzo, and B.~M.~Terhal, {\it Quantum Inf. Comput.} \textbf{7}, 297 (2007).

\bibitem{Stephens08}
A.~M.~Stephens, A.~G.~Fowler, and L.~C.~L.~Hollenberg, {\it Quantum Inf. Comput.} \textbf{8}, 330 (2008).

\bibitem{Stephens09}
A.~M.~Stephens and Z.~W.~E.~Evans, {\it Phys. Rev. A} \textbf{80}, 022313 (2009).

\bibitem{Spedalieri09}
F.~M.~Spedalieri and V.~P.~Roychowdhury, {\it Quantum Inf. Comput.} \textbf{9}, 666 (2009).

\bibitem{Stock09}
R.~Stock and D.~F.~V.~James, {\it Phys. Rev. Lett.} {\bf 102}, 170501 (2009).

\bibitem{Devitt09}
S.~J.~Devitt, A.~G.~Fowler, A.~M.~Stephens, A.~D.~Greentree, L.~C.~L.~Hollenberg, W.~J.~Munro, and K.~Nemoto, {\it New. J. Phys.} {\bf 11}, 083032 (2009).

\bibitem{Meter10}
R.~Van Meter, T.~D.~Ladd, A.~G.~Fowler, and Y.~Yamamoto, {\it Int. J. Quantum Inf.} {\bf 8}, 295 (2010).

\bibitem{Yao12}
N.~Y.~Yao, L.~Jiang, A.~V.~Gorshkov, P.~C.~Maurer, G.~Giedke, J.~I.~Cirac, and M.~D.~Lukin, {\it Nature Commun.} {\bf 3}, 800 (2012).
	
\bibitem{Nickerson1}
N.~H.~Nickerson, Y.~Li, and S.~C.~Benjamin, {\it Nat. Commun.} {\bf 4}, 1756 (2013).

\bibitem{Monroe12}
C.~Monroe, R.~Raussendorf, A.~Ruthven, K.~R.~Brown, P.~Maunz, L.-M.~Duan, and J.~Kim, arXiv:1208.0391.

\bibitem{Nemoto13}
K.~Nemoto, M.~Trupke, S.~J.~Devitt, A.~M.~Stephens, K.~Buczak, T.~N\"obauer, M.~S.~Everitt, J.~Schmiedmayer, and W.~J.~Munro, arXiv:1309.4277.

\bibitem{Barrett1}
S.~D.~Barrett and T.~M.~Stace, {\it Phys. Rev. Lett.} {\bf 105}, 200502 (2010).

\bibitem{Wang10}
D.~S.~Wang, A.~G.~Fowler, A.~M.~Stephens, and L.~C.~L.~Hollenberg, {\it Quantum Inf. Comput.} \textbf{10}, 456 (2010).

\bibitem{Wang11}
D.~S.~Wang, A.~G.~Fowler, and L.~C.~L.~Hollenberg, {\it Phys. Rev. A} \textbf{83}, 020302(R) (2011).

\bibitem{Fowler3}
A.~G.~Fowler, A.~C.~Whiteside, and L.~C.~L.~Hollenberg, {\it Phys. Rev. Lett.} \textbf{108}, 180501 (2012).

\bibitem{Devitt13}
S.~J.~Devitt, A.~M.~Stephens, W.~J.~Munro, and K.~Nemoto, {\it Nat. Commun.} {\bf 4}, 2524 (2013).

\bibitem{Mucciolo1}
E.~Novais and E.~R.~Mucciolo, {\it Phys. Rev. Lett.} \textbf{110}, 010502 (2013).

\bibitem{Mucciolo2}
P.~Jouzdani, E.~Novais, and E.~R.~Mucciolo, {\it Phys. Rev. A} \textbf{88}, 012336 (2013).

\bibitem{Ghosh1}
J.~Ghosh, A.~G.~Fowler, and M.~R.~Geller, {\it Phys. Rev. A} \textbf{86}, 062318 (2012).

\bibitem{Duclos-Cianci1}  	
G.~Duclos-Cianci and D.~Poulin, {\it Phys. Rev. Lett.} {\bf 104}, 050504 (2010).

\bibitem{Duclos-Cianci2}  		
G.~Duclos-Cianci and D.~Poulin, {\it Quantum Inf. Comput.} {\bf 14}, 721 (2014).

\bibitem{Bravyi1b}  
S.~Bravyi and J.~Haah, {\it Phys. Rev. Lett.} {\bf 111}, 200501 (2013).

\bibitem{Sarvepalli1}
P.~Sarvepalli and R.~Raussendorf, {\it Phys. Rev. A} {\bf 85}, 022317 (2012).

\bibitem{Wootton1} 
J.~R.~Wootton and D.~Loss, {\it Phys. Rev. Lett.} {\bf 109}, 160503 (2012).

\bibitem{Wootton2}
A.~Hutter, J.~R. Wootton, D.~Loss,  arXiv:1302.2669.

\bibitem{Wootton3}
J.~R.~Wootton, arXiv:1310.2393.

\bibitem{Delfosse1}
N.~Delfosse, {\it Phys. Rev. A} {\bf 89}, 012317 (2014).

\bibitem{Edmonds1}
J.~Edmonds, {\it Canad. J. Math.}, {\bf 17}, 449 (1965).

\bibitem{Bombin1}
H.~Bombin and M.~A.~Martin-Delgado, {\it Phys. Rev. Lett.} {\bf 97}, 180501 (2006).

\bibitem{Landahl1} 
A.~J.~Landahl, J.~T.~Anderson, and P.~R.~Rice, arXiv:1108.5738.

\bibitem{Gottesman97}
D.~Gottesman, Ph.D. thesis, California Institute of Technology, 1997.

\bibitem{Harrington04}
J.~Harrington, Ph.D. thesis, California Institute of Technology, 2004.

\bibitem{Wang03}
C.~Wang, J.~Harrington, and J.~Preskill, {\it Ann. Phys.} {\bf 303}, 31  (2003).

\bibitem{Kolmogorov1}
V.~Kolmogorov, {\it Math. Program. Comput.} {\bf 1}, 43 (2009).

\bibitem{Saito1}
M.~Saito and M.~Matsumoto, in {\it Monte Carlo and Quasi-Monte Carlo Methods 2006}, edited by A. Keller, S. Heinrich, and H. Niederreiter (Springer, Berlin, 2008), Vol. 2, p. 607.

\bibitem{BK05}
S.~Bravyi and A.~Kitaev, {\it Phys. Rev. A} {\bf 71}, 022316 (2005).

\bibitem{Stace1}
T.~M.~Stace and S.~D.~Barrett, {\it Phys. Rev. A} {\bf 81}, 022317 (2010).

\bibitem{Honecker}
A.~Honecker, M.~Picco, and P.~Pujol, {\it Phys. Rev. Lett.} {\bf 87}, 047201 (2001).

\bibitem{Merz}
F.~Merz and J.~T.~Chalker, {\it Phys. Rev. B} {\bf 65}, 054425 (2002).

\bibitem{Ohzeki09}
M.~Ohzeki, {\it Phys. Rev. E} {\bf 79}, 021129 (2009).

\bibitem{deQueiroz09}
S.~L.~A. de Queiroz, {\it Phys. Rev. B} {\bf 79}, 174408 (2009).

\bibitem{Ohno04}
T.~Ohno, G.~Arakawa, I.~Ichinose, and T.~Matsui, {\it Nucl. Phys. B} {\bf 697}, 462 (2004).

\bibitem{Poulin1}
D.~Poulin, {\it Phys. Rev. A} \textbf{74}, 052333 (2006).

\bibitem{Evans2}
Z.~W.~E.~Evans and A.~M.~Stephens, {\it Quantum Inf. Process.} \textbf{11}, 1511 (2012).

\bibitem{Bravyi13}
S.~Bravyi and A.~Vargo, {\it Phys. Rev. A} {\bf 88}, 062308 (2013).

\bibitem{Barrett2}
Y.~Li, S.~D.~Barrett, T.~M.~Stace, and S.~C.~Benjamin, {\it Phys. Rev. Lett.} {\bf 105}, 250502 (2010).

\bibitem{Fujii10}
K.~Fujii and Y.~Tokunaga, {\it Phys. Rev. Lett.} {\bf 105}, 250503 (2010).

\bibitem{Stephens13}
A.~M.~Stephens, W.~J.~Munro, and K.~Nemoto, {\it Phys. Rev. A} {\bf 88}, 060301(R) (2013).

\end{thebibliography}

\end{document}